\documentstyle[aps,prl,epsf]{revtex}
\def\beq{\begin{equation}}
\def\eeq{\end{equation}}
\begin{document}                
\title{An Experimentally Realizable Weiss Model for Disorder-Free Glassiness}
\author{P. Chandra,$^1$ M.V. Feigelman,$^2$  M.E. Gershenson$^3$ and L.B. Ioffe$^{2,3}$}
\address{$^1$NEC Research Institute, 4 Independence Way, Princeton NJ
08540}
\address{$^2$Landau Institute for Theoretical Physics, Moscow, RUSSIA}
\address{$^3$Department of Physics, Rutgers University, Piscataway, NJ 08855}
%
\maketitle
\begin{abstract}
We summarize recent work on a frustrated periodic long-range Josephson array
in a parameter regime where its dynamical behavior is {\sl identical} to
that of the $p=4$ disordered spherical model.  We also discuss
the physical requirements imposed by the theory on the experimental
realization of this superconducting network.
\end{abstract}
\vspace{0.5in}


The identification of the key features underlying the physics of
glass formation remains an outstanding problem of statistical
mechanics.
A glassy system has a ``memory'' of its past behavior and
does not explore all of phase space; it breaks ergodicity without
thermodynamic selection of a unique state
thereby defying description by the standard
Gibbs methods.
Though glass formation in the absence of intrinsic disorder is a
widespread
phenemenon, it remains poorly understood even at the mean-field level.
Because vitrification is a dynamical transition that is
not necessarily accompanied by a static one, it remains outside
the framework of conventional Landau theory.  The crucial
links between glasses with intrinsic and self-generated
disorder remain an area of active discussion.\cite{activity} Recently
several microscropic non-random models for glassiness
have been proposed; most were studied via mapping to
disordered systems.\cite{disorder}
What common/distinct features are characteristic of glasses with
and without randomness?  How similar/different are glasses
with short- vs. long-range interactions?
What aspects of existing theoretical models
are relevant for experiment?   
In this Proceeding we summarize our
current
efforts to address these questions in a periodic model that is
both analytically accessible\cite{Chandra95,Chandra96a}
and experimentally realizable.\cite{Gershenson96}  

\centerline{\epsfxsize=8cm \epsfbox{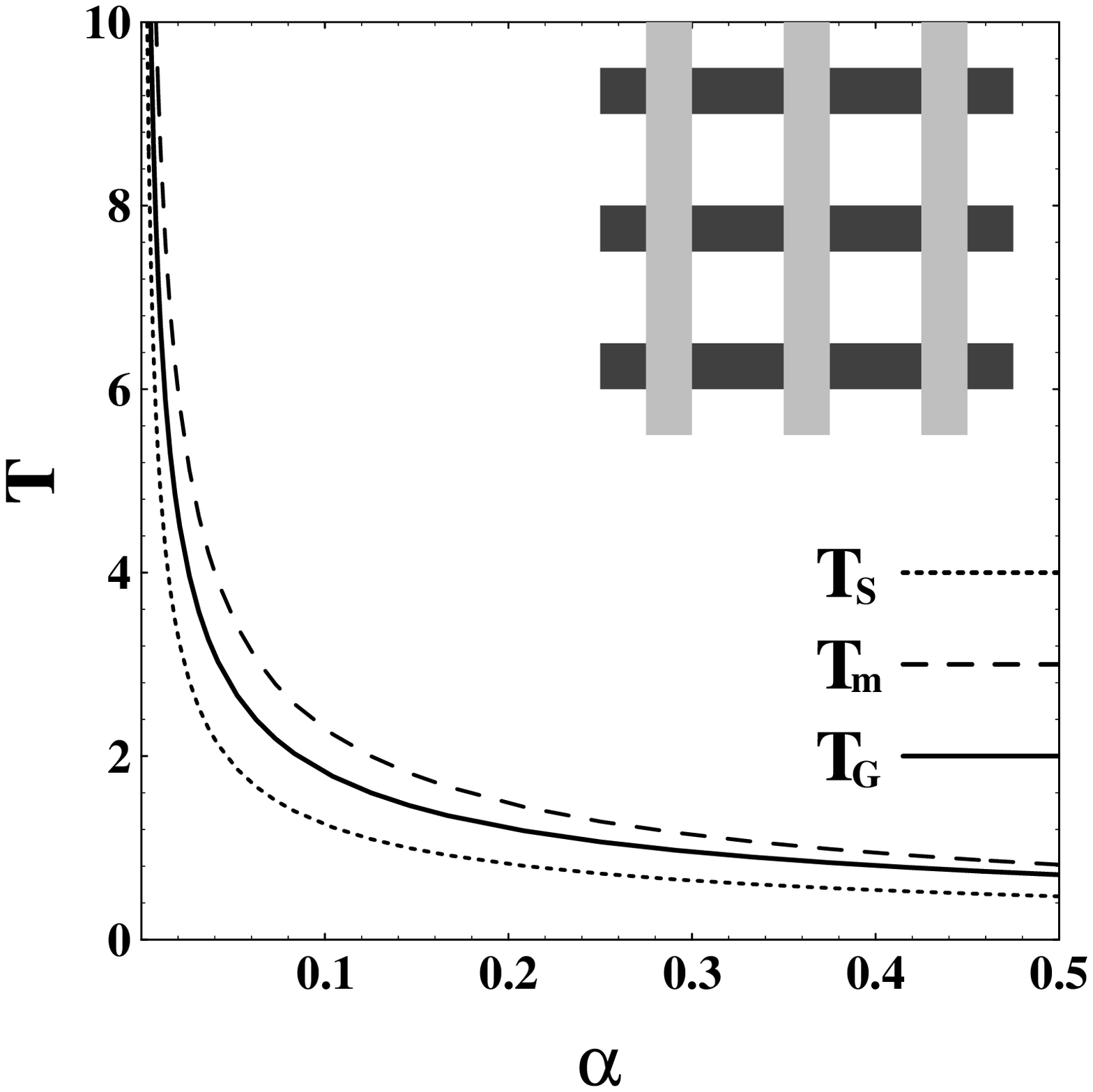}}
{\footnotesize {\bf Fig. 1} 
The phase diagram of the array (insert) where
$T_G$ indicates the temperature associated with the dynamical
instability
discussed in the text, $T_S$ is the speculated equilibrium
transition temperature and $T_m$ is the ``superheating'' temperature
where the low-temperature metastable states cease to exist.
}
\vspace{0.5in}

The physical system under study is a stack of two mutually perpendicular sets
of $N$ parallel thin
wires with Josephson junctions at each node (Figure 1)  that is
placed in an external tranverse field $H$.
The classical thermodynamic variables of this array are the $2N$
superconducting 
phases associated with each wire. In the absence of an external field
the
phase differences would be zero at each junction, but this
is not possible for finite $H$ and the phases are thus frustrated.
Here we assume that the Josephson couplings are sufficiently small so
that the induced fields are negligible in comparison with $H$; we
shall
return to this topic when discussing the experimental realization of
this network.
We can therefore describe the array by the Hamiltonian
\beq
{\cal H} = - \sum_{m,n}^{2N} s_m^{*} {\cal J}_{mn} s_n
\label{H}
\eeq
where ${\cal J}_{mn}$ is the coupling matrix
\beq
\hat{\cal  J} = \left( \begin{array}{cc}
0 & \hat{J} \\
\hat{J}^\dagger & 0
\end{array}
\right)
\label{J}
\eeq
with $J_{jk} = \frac{J_0}{\sqrt{N}} \exp(2\pi i \alpha jk /N)$ and $1 \! \leq
\! (j,k) \! \leq \! N$ where $j(k)$ is the index of the horizontal (vertical)
wires; 
$s_m = e^{i\phi_m}$  where the $\phi_m$ are the superconducting phases of the
$2N$ wires.
Here we have introduced the flux per unit strip, $\alpha = NHl^2/\phi_0$,
where $l$ is the inter-node spacing and $\phi_0$ is the flux quantum;
the normalization has been chosen so that $T_G$ does not scale with
$N$.

Because every horizontal (vertical) wire is linked to every vertical
(horizontal) wire, the number of nearest neighbors in this model is
$N$ and it is accessible to a mean-field
treatment.  For the same reason,
the free energy barriers separating its low-temperature solutions
scale with $N$. This situation is in marked contrast to that
in conventional 2D arrays where the coordination number and
hence the barriers are low.\cite{nets,Carini88}
A similar long-range network with positional disorder
was studied previously.\cite{Vinokur87} For $\alpha \gg {1/N}$ the
latter
system 
displays a spin glass transition which was mapped onto
the Sherrington-Kirkpatrick model\cite{Sherrington75} for $\alpha \gg 1$;
in this field regime there is no residual ``ferromagnetic''
phase coherence between wires.
Physically
this glassy behavior occurs because the phase differences associated with
the couplings $J_{jk}$ acquire random values and fill the interval
$(0,2\pi)$ uniformly for $\alpha \gg {1/N}$.  More specifically,
there will be no commensurability if the sum
\beq
\sum_{k=1}^N J_{jk}J^\dagger_{kl} = \sum_{k=1}^N \exp 2\pi i \left \{ {\alpha 
(j - l)}\over {N} \right \}k
\label{sum}
\eeq
is a smooth function of $(j - l)$ where $\{j,l\}$ ($k$) are indices labelling horizontal (vertical)
wires;
this will occur only if the expression in curly brackets on the
r.h.s. of
(\ref{sum}) is {\sl not} an integer for all $k$, a condition always
satisfied for the disordered array.
For the periodic case,
this situation is realized in the ``incommensurate window'' 
$1/N \ll \alpha \le 1$; here the phase-ordering unit cell is
larger than the system size so that the ``crystalline'' phase
is inaccessible.\cite{Chandra95} There are thus no special field values
where the number of low-temperature solutions are finite,
in contrast to the situation for $\alpha > 1$.

In the thermodynamic limit of $N \rightarrow \infty$ (with
fixed array area), the high-temperature approach to
the glass transition in the periodic array
has been studied\cite{Chandra95,Chandra96a}
using a modified Thouless-Anderson-Palmer (TAP) method.\cite{Thouless77}
Here we discuss the qualitative picture that emerges
from these results (cf. Fig. 1), referring the interested
reader elsewhere for a more detailed quantitative treatment.
\cite{Chandra95,Chandra96a}
As $T$ approaches $T_m^+$, where $T_m^+ \sim T_0 \approx {{J\sqrt{N}}\over
{2\sqrt {\alpha}}}$, there
appear a number of metastable states in addition to the paramagnetic
free-energy minimum. There is no dynamical transition at this
temperature, indicating that the system is not ``trapped''.
We speculate that this is because these states are 
energetically unfavorable and/or they
are separated by low barriers.
At lower temperatures, as $T \rightarrow T_G^+$, the paramagnetic
minimum 
is ``subdivided'' into an
extensive number of degenerate metastable states separated by effectively
infinite barriers, and the system is dynamically localized into one of them.
Qualitatively, in the interval $T_m > T > T_G$ there appear many
local minima in the vicinity of the paramagnetic state separated by 
{\sl finite} barriers; these barriers increase continuously and become
infinite at $T = T_G$.
Each of these minima is characterized by a finite ``site magnetization'' $m_i
= \langle s_i \rangle_T$  where``site'' refers to a wire.
When $T > T_G$ thermal fluctuations average over many states so that $\langle
m_i \rangle \equiv 0$.
At $T=T_G$ the system is localized in one metastable state and there is an
associated jump in the Edwards-Anderson order parameter, $\left (q =
\frac{1}{N}\sum_i \langle m_i\rangle^2 \right)$.
The low-temperature phase is characterized by a finite $q$ and by the presence
of a memory, $\lim_{t' \rightarrow \infty} \Delta(t,t') \neq 0$ where $\Delta
(t,t')$ is the anomalous response.
We expect that at $T = T_G$, the metastable states are degenerate and thus
there can be no thermodynamic selection.
However at lower temperatures interactions will probably break this degeneracy
and select a few states from this manifold;  then we expect an
($t \rightarrow \infty$) equilibrium first-order transition ($T_S$) which
should be accompanied by a jump in the local magnetization.
In order to observe this transition at $T_S$  the array must be
equilibrated
on a time-scale ($t_W$) longer than that ($t_A$) necessary to overcome
the barriers separating its metastable states; $t_A$  scales
exponentially with the number of wires in the array.
Thus the equilibrium transition at $T_S$ is observable {\sl only}
if $t_W \rightarrow \infty$ {\sl before} the thermodynamic limit
($N \rightarrow \infty$) is taken; in the opposite order of limits
only the dynamical transition occurs.

The periodic array thus exhibits a first-order thermodynamic     
transition {\sl preceeded} by a dynamical instability; the
glass transition at $T_G$ is characterized by a diverging
relaxation time and an accompanying jump in the
Edwards-Anderson order parameter.
Furthermore the dynamical coupled
equations for the response and correlation
functions of this {\sl non-random} network in the
regime ($1/N \ll \alpha < 1$) are {\sl identical} to
those obtained for the $p=4$ (disordered) 
spherical Potts model.\cite{Crisanti93}
By contrast the behavior of the disordered array is similar
to that of the $p=2$ case with coinciding static and dynamic transitions.
The possible connection between non-random glasses
and $p \ge 3$ (disordered) spherical models has been  previously
suggested
in the literature,\cite{Wolynes,Parisi,Franz} and the periodic
array results give support to those
conjectures.\cite{Chandra96a} Furthermore these long-range Josephson arrays
can be built in the laboratory, allowing for 
parallel theoretical and experimental studies of  the ``simplest spin
glasses''
that have previously been an abstraction.

Since the experimental realization of these arrays is important,
we would like to identify a number of simplifying assumptions
inherent in the theoretical treatment discussed here; in order
to test its predictions and probe beyond its realm (finite-size
effects etc.)
the fabricated system must satisfy
certain physical requirements.   In particular the model
has been only considered in the classical limit where quantum
mechanical fluctuations of the phase variables are negligible.
Furthermore the fields induced by the Josephson currents
must always be small compared to 
$H$ so that it can frustrate the phases effectively.
These constraints put some strong restrictions on the
choice of array parameters, which we elaborate below.

We begin with the first condition, that of maintaining
weak quantum fluctuations.
The strength of the latter is controlled by two parameters: 
the dimensionless normal resistance $e^2 R/\hbar$ and the ratio
$E_C/E_J$ where $E_C$ and $E_J$ are the Coulomb and the Josephson
energy scales respectively associated with each individual
phase variables.
The state of any array (or junction) is conveniently described by its position
on the plane  defined by these dimensionless parameters, $E_C/E_J$
and $e^2 R/\hbar$; the resulting plot is called a 
Schmid diagram.\cite{Schmid83}
For a conventional
short-range array $R \sim R_0$ and $E_C \sim e^2/C_0$ where $R_0$ 
($C_0$) is a resistance (capacitance) of an individual junction.
We require that the capacitance of the individual wires and the
dissipation in the system should be sufficiently large to
suppress the quantum fluctuations of phase and charge effects. 

According to the Ambegaokar-Baratoff relation,\cite{Ambegaoker63}
the normal-state resistance of a single junction in the
fabricated long-range arrays should be large ($R_0 \gg \hbar/e^2$) so that the
associated screening current is small;
for the sake of example we shall assume
$R_0 \approx 100\; kOhm$ and an individual capacitance of
$C \approx 1. \; 10^{-15}\; F$, parameters associated
with preliminary $Al-Al_2O_3-Al$ tunnel-junction arrays
already fabricated.\cite{Gershenson96} 
We note that 
the Josephson effect for a single tunnel-junction of such 
a large resistance (or for a short-range array of such junctions)  
would be   
completely overwhelmed by quantum
fluctuations and Coulomb blockade effects.
\cite{Delsing94,Iansiti88,Mooij90,Tinkham92}
Indeed, it is known that conventional nearest-neighbor
arrays  
exhibit insulating behavior if their individual
junction resistivitance is more than $15 kOhm$.\cite{Mooij90}
However in the long-range Josephson arrays all the junctions 
associated with
a particular wire are connected in parallel, so that the effective resistance 
and capacitance
determining their position on the Schmid diagram\cite{Schmid83}
are $R=R_0/N$ and 
$C=NC_0$ respectively.
Even for a small array with $N=100$  the effective resistance 
and capacitance are 
$R \sim 1\; kOhm$ and $C \sim 1. \; 10^{-13}\; F$
(corresponding to the charging energy $E_C \sim 10\; mK$) respectively
so that quantum fluctuations and charging effects are negligible
for these networks.
We note, however, that if the number of wires (and hence the
number of neighbors) were decreased significantly the parameters
of the individual junctions would have to be altered
to avoid complications of charging effects.
 
We would also like to minimize the effects of everpresent screening
currents to ensure that the external field frustrates the
superconducting
phases effectively.  More specifically we demand that the induced
flux in the array is less than a flux quantum
\beq
Li_{eff} < \Phi_O
\eeq
so that all fields and phase gradients produced by the diamagnetic
currents are negligible.  In order to determine the maximum induced
flux in the array we need only look at its largest loop; our condition
then becomes
\beq
(Nl)i_{eff} < \Phi_0
\eeq
where $l$ is the distance between nodes in the array.  For the
frustrated case, the maximum field occurs when the currents
add incoherently in one direction and coherently in the
other\cite{Vinokur87} so that
\beq
i_{eff} = \left ( \sqrt{N} N \right ) i_c.
\eeq
where $i_c$ is the critical current associated with an
individual junction.
Using 
\beq
i_c \sim  {{2e}\over {\hbar}}  E_J = {{2e}\over {\hbar}} \left( {{k_B T_c}\over
{\sqrt{N}}}\right)
\eeq
we find that the necessary
condition for weak induced fields becomes
\beq
E_J \lesssim \frac{\Phi_0^2}{l N^{5/2}}.
\label{E_J}
\eeq
Thus there is a delicate balance to be obtained between
the need for many neighbors (and high free-energy barriers)
and the requirement (\ref{E_J}) that the effective London
penetration length be substantial larger than the system size;
unfortunately early experimental attempts to realize these arrays
were not able to satisfy these restrictions due to resolution
limitations associated with the optical lithography methods
used.\cite{Sohn93a,Sohn93b}
Furthermore the predicted glassy behavior 
is only observable in the 
field range $\phi_0/(N l)^2 < H < \phi_0/l^2 $;
where the minimum and maximum correspond to a single flux quantum
in the whole array and in a single plaquette.
Even with the use of the e-beam lithography ($l = 0.5\; \mu m$) 
the requirement 
(\ref{E_J}) can only be satisfied for an array of $Al-Al_2O_3-Al$ junctions
($E_J = 1. \;10^{-17}\; Erg$) with the number of wires $N$ less than $300$.  
For these parameters with $N=100$
$\phi_0/(N l)^2$ is of the order of $0.01 \;Oe$, and a
careful screening of the earth's magnetic field is necessary. 
 
Once the arrays are fabricated to specification what measurements
should be performed?  The field-dependence of the critical
temperature of both disordered and periodic arrays is
crucial in establishing the effectiveness of
of the external field in frustrating the phases.
It is also a good test of ``phase freezing'' into a glassy
phase, and 
the functional dependence $T_c(H) \propto H^{-1/2}$ is expected
over a wide field and temperature range for both types of arrays.  
In order to firmly establish the presence of large barriers
imperative for glassiness, the history-dependence of the
critical current $j_c$ should be studied (i.e. its dependence
on the {\sl path} in the $T-H$ plane) in both arrays.
We note that below $T_c(\alpha)$ we expect a faster increase
in $j_c(T)$ for the periodic network than for its disordered
counterpart;\cite{Vinokur87} specifics for the former case
have yet to be determined.
We
expect that the diverging
relaxation time at $T_c$ can be accessed experimentally
via the a.c. response to a time-varying magnetic field
$H(t)$;
the associated ac susceptibility is
\beq
\chi_\omega = {{\partial M_\omega}\over {\partial H_\omega}} =
{{{C(\alpha,N)\omega}\over {\omega + i/t_R(T)}}}
\label{chiw}
\eeq
where $t_R$ is the longest response of the system that
diverges at $T = T_G$
and $C(\alpha,N)$ is to be found elsewhere.\cite{Chandra96a}
The $\omega \rightarrow 0$
limit of the a.c. susceptibility jumps to a finite
value at $T=T_G$, indicating the development of a finite
superconducting stiffness at the transition. Therefore
measurement of this a.c. response in a fabricated array would probe
its predicted glassiness.

Work is currently in progress to study the physical properties
of the periodic model in its non-ergodic regime ($T < T_G$).\cite{Chandra96b} 
We note that the glass transition temperature $T_G$
described above corresponds to the system going out of
equilibrium as it is cooled infinitesmally slowly
from high temperatures.  In practice there will always
be a finite-rate of cooling and the effective
glass transition will occur when the system drops
out of equilibrium.
If we define the reduced temperature $\Theta \equiv \left( {{(T -
T_G)}\over {T_G}} \right)$ then the effective glass transition 
will occur when the time associated with cooling, $t_c \approx 
{{\Theta} \over {d\Theta/dt}}$
is equal
to the relaxation time at that temperature, $t_R \approx t_0 \theta^{-\nu}$
where $t_0$ and $\nu$ are determined from the high-temperature
analysis elsewhere;\cite{Chandra96a}
the effective glass transition temperature and
the associated time are indicated as $\tilde{T_G}$ and $\tilde{t_G}$
in Figure 2.
There is evidence for history-dependence; the system's
response is very different if an additional field
is turned on at a time, $t_H$, during or after the cooling process
and a subsequent measurement is taken at time $t_M$ (cf. Figure 2),
reminiscent of the zero-field cooled vs. field-cooled susceptibility
observed in spin glasses. Furthermore an ``ageing'' time-scale
$t^*$ seems to exist in this system (cf. Fig. 2); 
if the field is turned
on during the cooling process at time $t_H$ and a subsequent
 measurement is taken
at time $t_M$ such that $t_H - t_M < t^*$ the system ``remembers'' the
cooling process;  otherwise ($t_H - t_M > t^*$) it ``forgets'' it
completely. Preliminary results suggest that $t^*$ has a decreasing
temperature-dependence and probably vanishes at low temperatures.
This presence of an additional ``ageing'' time-scale does not
exist in the Sherrington-Kirkpatrick theory, though it is indeed
observed in experimental spin glasses; it is thus amusing
that it emerges from the study of a non-random model.

\centerline{\epsfxsize=8cm \epsfbox{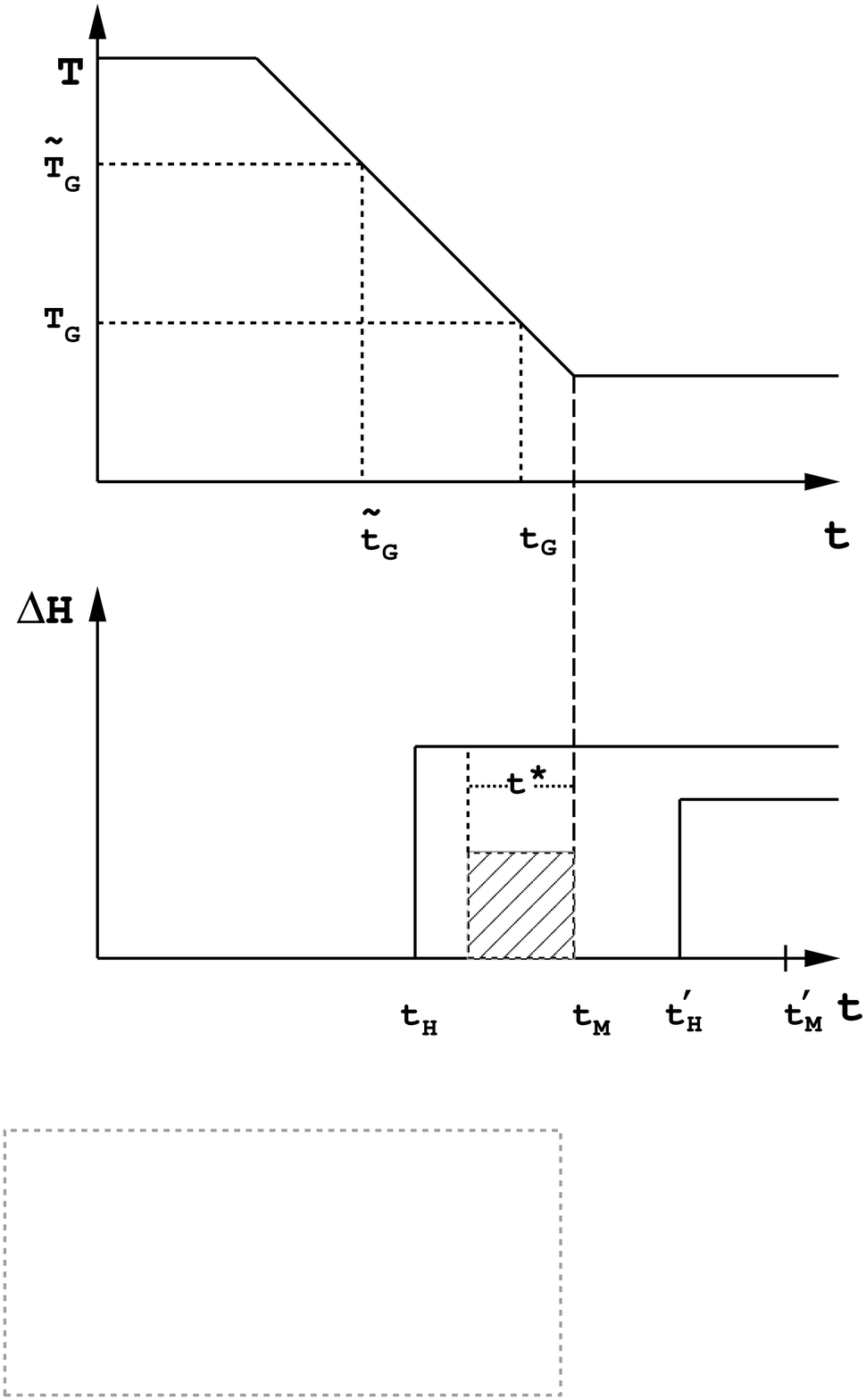}}
{\footnotesize {\bf Fig. 2}
Schematics of (a) Temperature ($T$)
(b) Application of an Additional Field   ($\Delta H$)vs time ($t$)
indicating the time-scales
involved in a finite-cooling experiment as discussed in
the text; $\tilde{T_G}$ and $\tilde{t_G}$ refer to the ``effective
glass transition'' when the system goes out of equilibrium, 
and $t_H$, $t_M$ and $t^*$
are the time-scales associated with the onset of an additional
field, a subseqent measurement and ageing.
}
\vspace{0.5in}

In conclusion we have presented a summary of recent work on a periodic
model that displays ``freezing'' into one of an extensive
number of metastable states without thermodynamic selection.
This glass transition is characterized by a diverging relaxation
time and an accompanying jump in the Edwards-Anderson order parameter.
At low field strengths the dynamical behavior of this system
is identical to that of the $p=4$ (disordered) spherical model.
A key advantage of this system is that it can be built in the
laboratory, thereby allowing for the possibility of detailed
interplay between theory and experiment.  The behavior of
a similar array with a reduced number of neighbors is a completely
open question that could be important in understanding the
relevance of long-range models to real experimental glasses.
Finally, since any uncertainty in the position of the wires
introduces randomness in the array, the continued study of
this network could also offer an opportunity to study the
crossover between glasses with spontaneously generated and
quenched disorder.


\begin{references}

\bibitem{activity} e.g. T.R. Kirkpatrick and D. Thirumalai, PRB {\bf 36}, 
5388 (1987); 
J.P. Bouchaud
and M. Mezard, J. Phys. I {\bf 4}, 1109 (1994); 
S. Franz and J. Herz, PRL {\bf 74}, 2115 (1995); M. Potters and G. Parisi,
unpublished (cond-mat preprint 9503009) and references therein.

\bibitem{disorder} e.g. E. Marinari, G. Parisi and F. Ritort, {\sl J. Phys. A},
{\bf 27}, 7647 (1994); M. Mezard and G. Parisi, unpublished.

\bibitem{Chandra95} P. Chandra, L. B. Ioffe and D. Sherrington, Phys. Rev.
Lett. {\bf 75}, 713 (1995).

\bibitem{Chandra96a} P.Chandra, M.V. Feigelman and  L.B. Ioffe, 
Phys. Rev. Lett. {\bf 76}, 4805 (1996).

\bibitem{Gershenson96}M.E. Gershenson et al., To Be Published.

\bibitem{nets}e.g. J.E. Mooij and G.B.J. Sch\"on ed., ``Coherence in 
Superconducting Networks,'' Physica {\bf 152B}, 1 (1988).

\bibitem{Carini88} J. P. Carini, Phys. Rev B {\bf 38}, 63 (1988). 


\bibitem{Vinokur87} V. M. Vinokur, L. B. Ioffe, A. I. Larkin, M. V. Feigelman,
Sov. Phys. JETP {\bf 66}, 198 (1987).

\bibitem{Sherrington75} D. Sherrington and S. Kirkpatrick, Phys. Rev. B {\bf
35}, 1792 (1975).

\bibitem{Thouless77}D.J. Thouless, P.W. Anderson and R.G. Palmer, {\sl
Phil. Mag.}, {\bf 35}, 593 (1977).

\bibitem{Crisanti93} A. Crisanti, H. Horner, H.-J. Sommers, 
Z. Phys. B, {\bf 92}, 257 (1993).

\bibitem{Wolynes}T.R. Kirkpatrick and D. Thirumalai, {\sl
Phys. Rev. B},
{\bf 36}, 5388 (1987); T.R. Kirkpatrick, D. Thirumalai and
P.G. Wolynes,
{\sl Phys. Rev. B}, {\bf 40}, 1045 (1989).

\bibitem{Parisi} G. Parisi in J.J. Brey, J. Marro, J.M. Rubi
and M. San Miguel, eds. {\sl Twenty Five Years of Non-Equilibrium
Statistical
Mechanics; Proc. of the Thirteenth Sitges Conference},
(Springer-Verlag, Berlin 1995), pp. 135-42.

\bibitem{Franz}S. Franz and J. Herz, {\sl Phys. Rev. Lett.}, {\bf 74},
2115 (1995(.

\bibitem{Schmid83} A. Schmid, Phys. Rev. Lett. {\bf 51}, 1506 (1983);
R. Fazio and G. Shon, Phys. Rev. B {\bf 43}, 5307 (1991).

\bibitem{Ambegaoker63} V. Ambegaoker and A.Baratoff, Phys. Rev. Lett. \underline{10},
486 (1963).

\bibitem{Delsing94} P. Delsing, C. D. Chen, D. B. Haviland, Y. Harada and T.
Cleson, Phys. Rev. B {\bf 50}, 3959 (1994); J. E. Mooij and G. Shon in ``Single
Charge Tunneling: Coulomb Blocade Phenomena in Nanostructures'' ed. by H.
Grabert and M. Devoret, Plenum NY (1992).

\bibitem{Iansiti88}M.Iansiti, A.T.Johnson, C.J.Lobb, and M.Tinkham. Phys. Rev. Lett. {\bf 60}, 
2414 (1988).

\bibitem{Mooij90} J. E. Mooij, B. J. van Wess, L. J. Geerligs, M. Peters, R.
Fazio and G. Shon, Phys. Rev. Lett. {\bf 65}, 645 (1990).


\bibitem{Tinkham92} M. Tinkham in in ``Single Charge Tunneling: Coulom Blocade
Phenomena in Nanostructures'' ed. by H. Grabert and M. Devoret, Plenum NY
(1992). 

\bibitem{Sohn93a}
L. L. Sohn, M. T. Tuominen, M. S. Rzchowski, J. U. Free, and M. Tinkham.
Phys.Rev.B 47, 975 (1993).  

\bibitem{Sohn93b}
L. L. Sohn, M. S. Rzchowski, J.U.Free, and M.Tinkham.
Phys. Rev. B {\bf 47}, 967 (1993). 

\bibitem{Chandra96b} P. Chandra, M.V. Feigelman, L.B. Ioffe and D.M. Kagan,
To be published. 


\end{references}
\end{document}